# Mechanical Properties of Ultralow Density Graphene Oxide/Polydimethylsiloxane Foams

Cristiano F. Woellner[1,2], Peter S. Owuor[3], Tong Li[2], Soumya Vinod, Sehmus Ozden[2], Suppanat Kosolwattana[2], Sanjit Bhowmick[5], Luong X. Duy[3], Rodrigo V. Salvatierra[3], Bingqing Wei[5], Syed A. S. Asif[5], James M. Tour[3], Robert Vajtai[3], Jun Lou[3], Douglas S. Galvão[1,2], Chandra S. Tiwary[3], Pulickel. M. Ajayan[3]

[1]*Applied Physics Department, State University of Campinas, 13083-859 Campinas-SP, Brazil*

[2]*Center for Computational Engineering & Sciences, State University of Campinas, Campinas-SP, Brazil*

[3]*Department of MSNE Rice University Houston, TX 77005, USA*

[4]*Mechanical Engineering University of Delaware DE 19717, USA,*

[5]*Hysitron, Inc. Minneapolis, MN 55344, USA*

ABSTRACT

*Low-density, highly porous graphene/graphene oxide (GO) based-foams have shown high performance in energy absorption applications, even under high compressive deformations. In general, foams are very effective as energy dissipative materials and have been widely used in many areas such as automotive, aerospace and biomedical industries. In the case of graphene-based foams, the good mechanical properties are mainly attributed to the intrinsic graphene and/or GO electronic and mechanical properties. Despite the attractive physical properties of graphene/GO based-foams, their structural and thermal stabilities are still a problem for some applications. For instance, they are easily degraded when placed in flowing solutions, either by the collapsing of their layers or just by structural disintegration into small pieces. Recently, a new and scalable synthetic approach to produce low-density 3D macroscopic GO structure interconnected with polydimethylsiloxane (PDMS) polymeric chains (pGO) was proposed. A controlled amount of PDMS is infused into the freeze-dried foam resulting into a very rigid structure with improved mechanical properties, such as tensile plasticity and toughness. The PDMS wets the graphene oxide sheets and acts like a glue bonding PDMS and GO sheets. In order to obtain further insights on mechanisms behind the enhanced mechanical pGO response we carried out fully atomistic molecular dynamics (MD) simulations. Based on MD results, we build up a structural model that can explain the experimentally observed mechanical behavior.*

# INTRODUCTION

Low-density porous foam materials based on graphene/graphene oxide (GO) structures are very promising to energy absorption applications, even under high compressive deformations [1-4]. Structural foams are widely used in packing industry to cushion delicate products and also as weight saving materials. Other areas have also benefited from these 3D foams, such as biomedical [5], fuel cells [6], supercapacitors [7], flexible electronics [8-9], among others. The recently proposed freeze drying method to produce PDMS-graphene/GO (pGO) foams [3] has been proved to be an elegant and cost effective way to produce these foams. Despite the attractive physical properties of these foams, their stability still needs improvement. For instance, they are easily degraded when placed in solution, either by the collapse of their layers or just structural disintegration into small pieces. Bending and buckling of the layers are main causes of failure when subjected to compression loads [11]. Low bending modulus of the individual nanosheets in graphene/GO has limited their use as effective load carrying materials. Also, individual nanosheets are normally held together by week van der Waal forces [12], which results in very poor shear strength. Recently, several reports have shown that connecting these nanosheets, either by polymer or carbon nanotubes (CNTs) infiltration and/or welding, can result into foams with very high stiffness [10,13]. However, little is known about the failure mechanisms of nanostructured 3D freeze-dried foams with a controlled polymer infiltration. A better understanding of these mechanisms is important to produce foams with enhanced mechanical properties, in particular with relation to their mechanical stability without compromising their high strength and low-density values.

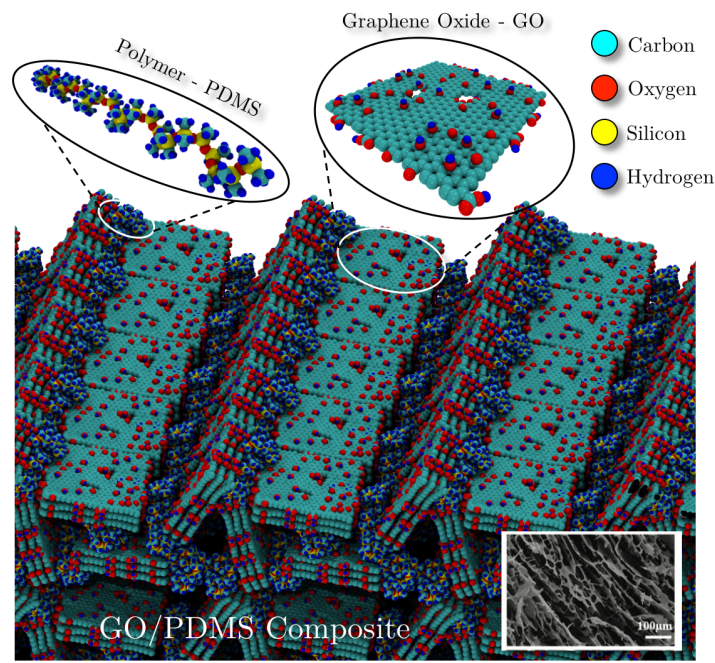

**Figure 1.** Schematics of the proposed structural PDMS-GO (pGO) model and its components, see details in [3].

In this work we have investigated the origin of the reported pGO mechanical enhanced properties (see Figure 1) through fully atomistic molecular dynamics (MD) simulations.

**THEORY AND SIMULATION DETAILS**

All MD simulations were carried out using the ReaxFF force field [14], as implemented in the open source code LAMMPS [15]. ReaxFF is used in classical MD simulations but with the advantage of accurately describing chemical processes, such as formation and breaking of covalent bonds. Its parameterization is obtained using Density Functional Theory (DFT) calculations and its accuracy, compared with experimental data, is around 2.9 kcal/mol for unconjugated and conjugated systems, respectively [14]. This allows ReaxFF to handle large systems, unlike fully quantum methods. All structures were generated using VMD/TopoTools [16]

Our structural model systems are composed of GO square sheets of 41 Å in length and containing 704 carbon, 91 oxygen, and 42 hydrogen atoms. The oxygen/carbon ratio chosen in the model was 13% and the functional group distribution, in terms of the number of oxygen, was given by 31% for carboxyl groups (all in the edges), 31% for hydroxyl groups and 38% for epoxy groups (14% on the edges) [3,10]. All functional groups were randomly distributed along the graphene sheet. The resulting GO sheet was kept the same for all MD simulations. The PDMS chains were composed of 30 repeat units (monomers), containing a total of 60 carbon, 182 hydrogen, 30 oxygen, and 30 silicon atoms. Throughout the MD simulations we kept the mass ratio (polymer/GO) around 10%. To describe the PDMS polymer and its interaction with the GO sheets we used the force field developed by Chenoweth *et al.*, described in detail in ref. [17].

To simulate the stress–strain process the entire system was thermalized for 600 ps at 300 K, followed by the application of a time dependent external force on each carbon atom in the GO sheets. This force is applied along opposite directions in order to pull the sheets apart (see Figures 2 and 3). The applied force is maintained until the sheets breaks apart. In this study, three different arrangements of GO sheets covered with PDMS chains were considered, as shown in Figure 3.

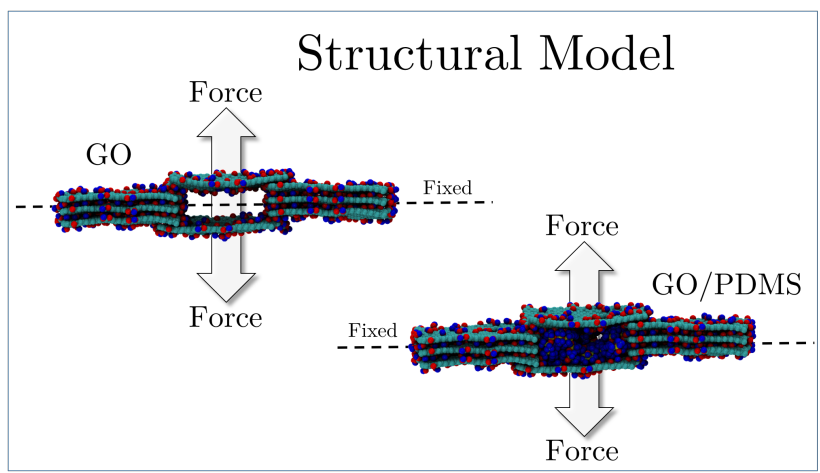

**Figure 2**. Structural models used to describe the mechanical response with/without PDMS in GO foams.

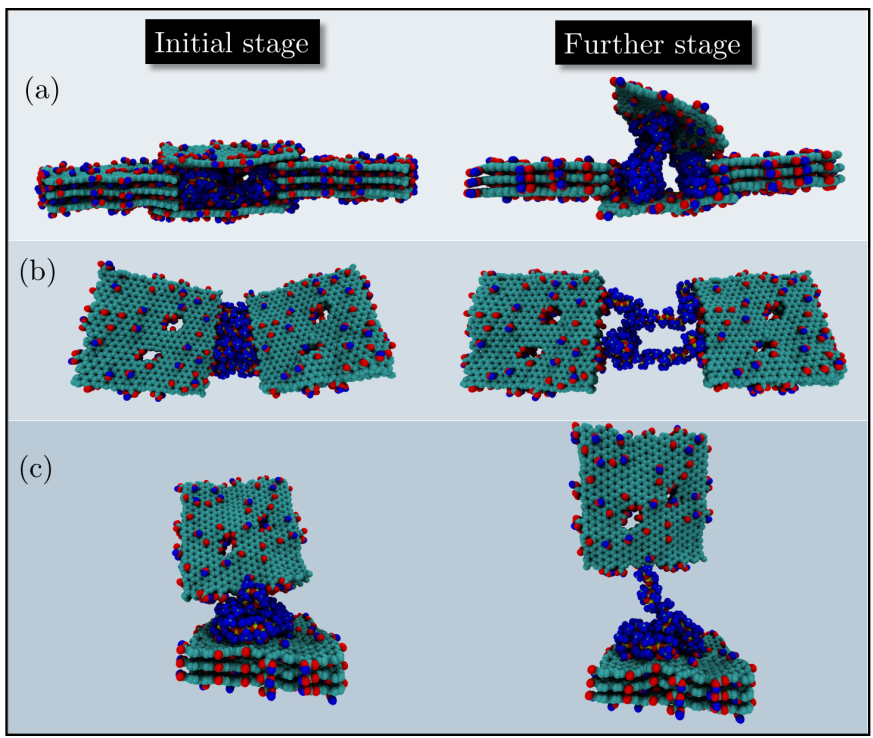

**Figure 3**. (Left) Schematic of pGO structural models for three different considered arrangements. (Right) MD snapshots at later stages, where we can already see the layer separations.

**RESULTS AND DISCUSSION**

As mentioned above, it was experimentally observed [3] that the controlled addition of a thin PDMS layer into GO materials can form an interconnected foam structure (pGO - Figure 1). PDMS was chosen due to its high affinity to GO nanosheets, with both physical adsorption and chemically covalent cross-links. Detailed mechanical characterization under tensile and compressive loading was performed and a significant enhancement of the mechanical properties was observed [3].

In order to gain further insights into the influence of the PDMS in the mechanical properties of the pGO foam, we created a simplified structural model shown in the Figures 1-3. To investigate the stress–strain behavior of the interconnected foam, we considered three different arrangements of GO sheets interconnected with PDMS, see Figure 3. The orientations are: (a) two neighboring three-layers GO anchoring two single layers GO sheets connected with PDMS; (b) two neighboring three-layers GO with their relative edges facing each order and with PDMS in between, and; (c) rotated by 90°. These configurations were chosen based on structural information gained from electron microscopy [3].

In Figure 3 we present representative MD snapshots of this process. Each snapshot in the initial stage (left side of Figure 3) represents the thermalized structure (at 300 K) obtained from reactive molecular dynamics (MD) simulations. The further stage snapshots (right side of Figure 3) were taken from stress–strain MD simulations before structural failure occurs. From this Figure we can see ductile-like pGO deformations, in contrast to what happens with pure GO, which breaks into pieces exhibiting a pronounced brittle behavior. This increased fracture resistance can be attributed to the presence of long PDMS chains, which clearly show their capability of holding GO sheets together up to remarkable strain values.

In Figure 4 we present the relative displacement as a function of the applied force on the outer GO sheets. This arrangement was considered in order to provide a reliable comparison (in terms of the energies involved) between pGO and GO foams. From the figure, we can clearly see that the necessary force to pull the GO sheets apart is larger in the case where the foams are connected with PDMS, which is consistent with pGO being more resilient to deformations than pure GO. This can be explained by the high affinity between PDMS and GO nanosheets, which is consistent with the experimental observations [3].

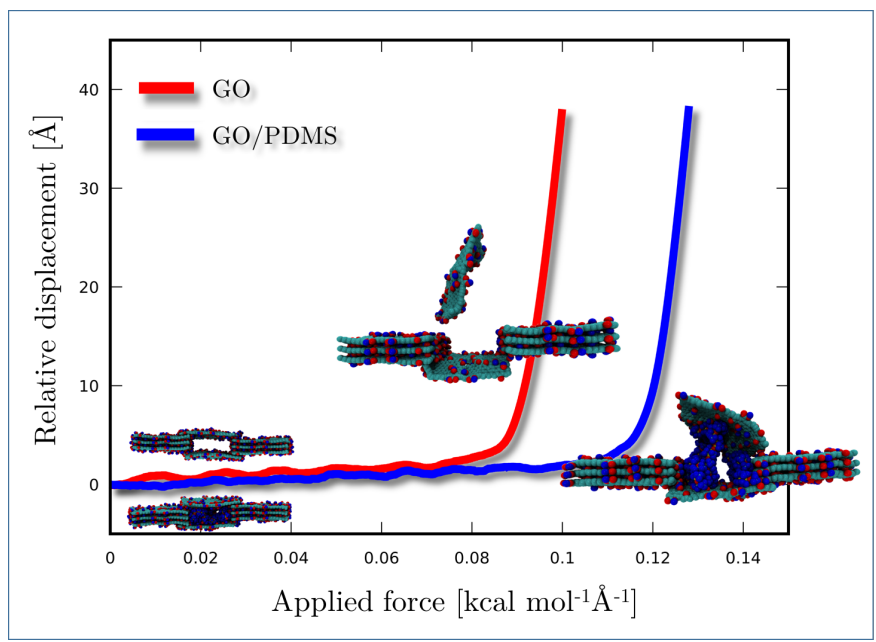

**Figure 4**. Relative displacement as a function of the applied force on the outer GO sheets.

**CONCLUSION**

We have investigated through fully atomistic molecular dynamics (MD) simulations the mechanisms of the experimentally observed enhanced mechanical properties of polydimethylsiloxane (PDMS)-graphene oxide foams (pGO) [3]. PDMS wets the individual GO sheets thereby preventing premature collapsing under load. As a result of PDMS infusion into the GO foams, the resulting interconnected pGO exhibits a significant improvement of their mechanical properties. Fracture mechanisms of pGO were also investigated by fully atomistic MD simulations and the results can explain the experimentally observed foam enhanced mechanical properties [3].

ACKNOWLEDGMENTS

The authors thank the Air Force Office of Scientific Research (Grants FA9550-13-1-0084 and MURI FA9550-12-1-0035) for funding this research. CFW acknowledges the São Paulo Research Foundation (FAPESP) Grant No. 2014/24547-1 for financial support. Computational and financial support from the Center for Computational Engineering and Sciences at Unicamp through the FAPESP/CEPID Grant No. 2013/08293-7 is also acknowledged.